# A Generative AI-driven Metadata Modelling Approach

Mayukh Bagchi

DISI, University of Trento, Italy.
Institute for Globally Distributed Open Research and Education (IGDORE).

mayukh.bagchi@igdore.org

**Abstract**
Since decades, the modelling of *metadata* has been core to the functioning of any academic library. Its importance has only enhanced with the increasing pervasiveness of Generative Artificial Intelligence (AI)-driven information activities and services which constitute a library's outreach. However, with the rising importance of metadata, there arose several outstanding problems with the process of designing a library metadata model impacting its reusability, crosswalk and interoperability with other metadata models. This paper posits that the above problems stem from an underlying thesis that there should only be a *few* core metadata models which would be *necessary and sufficient* for any information service using them, irrespective of the heterogeneity of intra-domain or inter-domain settings. To that end, this paper advances a contrary view of the above thesis and substantiates its argument in three key steps. First, it introduces a *novel* way of thinking about a library metadata model as an ontology-driven composition of five functionally interlinked *representation levels* from perception to its intensional definition via properties. Second, it introduces the representational manifoldness implicit in each of the five levels which cumulatively contributes to a *conceptually entangled* library metadata model. Finally, and most importantly, it proposes a *Generative AI-driven Human-Large Language Model (LLM) collaboration* based metadata modelling approach to disentangle the entanglement inherent in each representation level leading to the generation of a *conceptually disentangled* metadata model. Throughout the paper, the arguments are exemplified by motivating scenarios and examples from representative libraries handling cancer information.

## Introduction

Since the 1980s (Yu and Breivold 2008) and increasingly from the 2000s (Tedd and Large 2004), the development and continual management of (digitised) *metadata* (Satija, Bagchi and Martínez-Ávila 2020) has been core to the mundane functioning of any well-administered academic library. With the advent and increasing pervasiveness of AI (Cox, Pinfield and Rutter 2019) and *Generative AI* (Banh and Strobel 2023), however, the scope of an academic library has radically expanded to include the deployment of AI-based data-driven information services, e.g., *research data* management (Andrikopoulou, Rowley and Walton 2022), *ontology*-driven content management (Bagchi 2021a; Bagchi 2021b), *chatbots* (Bagchi 2020), all of which crucially depend on a *well-founded* metadata model (semantically) annotating and exposing the underlying data. For instance, let us consider the motivating scenario of semantically annotating scientific information in cancer research (e.g., cancer big data (Jiang *et al*. 2022)) within the framework of an academic library. Such an exercise is clearly *dependent* on a multitude of factors, e.g., the precise purpose, the target user base, the technical features, etc., each of which mutually defines the information service(s) that will harness such information. To that end, for example, a metadata model designed by a metadata librarian (Han and Hswe 2011) annotating, e.g., cancer big data, for a specialised cancer research library will be *significantly* (if not completely) different from one suited for an oncology policy library which, again, would be considerably different from a metadata model employed by a medical college library.

The general thesis advanced by this paper, as evidenced from the aforementioned motivating scenario and several other similar use-cases (see, for e.g., (Van Dijck 2014; Gartner 2016; Ulrich *et al*. 2022)), is that *there is no unique metadata model which is necessary and sufficient* for semantic data annotation within a single domain and certainly (not) for the heterogeneity inherent in cross-domain use-case scenarios (see also (EU-NSF 2024)). Let us understand, via the lens of the motivating scenario, the *interlinked representation levels* which cumulatively constitute a metadata model as indicated in the above thesis. First, the *perception* of the (expert) users of a cancer research library towards cancer research data would be highly specialised (e.g., uncovering multi-omics workflows (Ulrich *et al*. 2022)) and hence considerably different to how it is perceived by the (expert) users of an oncology policy library or a medical college library. Second, as a partial consequence of the first reason, the *terminology* employed by the users of the three types of libraries to describe the various perceived concepts of cancer research data would be mutually different. Third, the decision to *ontologically* characterise a perceived data concept as, e.g., a function or a process (Arp and Smith 2008) or an object property or a data property (Bagchi 2022) etc., by users of the three different types of libraries would be mutually different. Fourth, as a direct consequence of the ontological characterization, the *taxonomy* (Bagchi and Madalli 2019) assumed by the users of the three types of libraries would be markedly different. Fifth, the *intensional* characterization (see, e.g., (Von Fintel and Heim 2011), for an overview of intensional semantics) of the taxonomy in terms of interrelating and describing its constituent concepts via object properties and data properties would be different for different sets of (expert) library users. Finally, it is also key to note that the ordered representation levels, as briefed above, (each) independently as well as cumulatively *compounds* the design of any metadata model (e.g., developed by one of the three aforementioned libraries) and complicates its *crosswalk* with any other metadata model (e.g., developed by the other two).

Notice that the above contextualization of the motivating scenario into a general thesis about modelling metadata is *not* by chance. It is, in fact, a direct instantiation of the problem of *Conceptual Entanglement* (Bagchi and Das 2022; Bagchi and Das 2023) ubiquitous in knowledge organization and representation research, whereby, it posits that any conceptual knowledge model (e.g., a metadata model) is *representationally manifold* by design and *cannot be* necessary and sufficient for all use-cases irrespective of their intra-domain or inter-domain setting. Let us understand the above in terms of the motivating scenario. First, there is always a *many-to-many* correspondence (hereafter, referred to as *representational manifoldness*) between entities and how they are perceived as concepts (e.g., by sets of users of the three different libraries). Second, given perception, there is always a representational manifoldness between the perceived concepts and how they are linguistically labelled (e.g., by the same sets of users) using some terminology. Third, given labelling, there is always a representational manifoldness between the labelled concepts and their ontological status (e.g., by the same sets of users). Fourth, as a consequence of the manifoldness in ontological characterisation, there is always a representational manifoldness between the ontologically characterised concepts and how they ought to be taxonomically classified (e.g., as per the warrant of the same set of users). Fifth, there is always a representational manifoldness between the taxonomically classified concepts and how they are intensionally interrelated and described via properties. Two observations. Firstly, it is interesting to note how the individual as well as the cumulative impact of the representation layers *magnify* the entanglement in the final metadata model. Secondly, and perhaps most importantly, notice the need to *explicate the decision of the modeller* (e.g., a metadata librarian) at each level which, in an overwhelming majority of cases, remains *implicit* (Bagchi and Das 2022; Bagchi and Das 2023).

The solution proposed in this paper is a *Generative AI-driven LLM-based* enhancement of an early version of the approach termed as *Conceptual Disentanglement* (Bagchi and Das 2022; Bagchi and Das 2023). The key focus of the approach is to *explicitly disentangle* the decisions made by the modeller (e.g., a metadata librarian), at each (knowledge) representation level mentioned before, which would *otherwise implicitly entangle* the final conceptual knowledge artefact (e.g., the metadata model). To that end, the general strategy of the approach is to explicitly enforce *one-to-one* correspondence (hereafter, referred to as *representational bijection*) out of the potentially many representational manifold possibilities at each level. To explain using the motivating scenario, first, the metadata librarian (who is the modeller) should explicate a representational bijection between entities and their consensual perception as concepts (e.g., by a particular set of users of one of the libraries). Second, given the fixation of perception, the metadata librarian should explicate a representational bijection between the perceived concepts and their linguistic labelling via a consensual body of terminology. Third, given the fixation of terminology, the metadata librarian should explicate a representational bijection between the labelled concepts and their ontological commitments (Guarino, Carrara and Giaretta 1994), i.e., whether they are events or processes or properties, etc. In fact, an ontology-driven metadata model is *key to underpin the precise semantics* represented within (meta)data concepts and properties (see, for e.g., the arguments advanced in (Dutta 2014; Leipzig *et al*. 2021)). Fourth, the metadata librarian should explicate a representational bijection between the ontologically characterised concepts and their exact taxonomy. Fifth, given the taxonomy, the metadata librarian should explicate a representational bijection between each concept in the taxonomy and their exact intensional characterization. Finally, it is interesting to note how the individual as well as the cumulative disentanglement of the representation layers *minimise the entanglement* in the final metadata model. Notice that the decision of the modeller is also *explicit* at each representation level thereby crucially impacting the minimization of the conceptual entanglement.

While the contextualization, problem and solution approach is clear to some extent, it is not yet clear as to how the solution approach can be potentially *implemented* by, e.g., a *metadata librarian*, within an academic library setting. In fact, there are two critical complementary highlights which should characterise any implementation strategy of the conceptual disentanglement approach for modelling library metadata. First, while the approach attempts to disentangle, level-by-level, the entanglement inherent by design in a metadata model, it runs the risk of considerable intellectual work on part of the metadata librarian in terms of manually factoring and refactoring constituent knowledge management nitty-gritties (Bagchi 2019a; Bagchi 2019b). On the other hand, given the justified complexity of the conceptual disentanglement approach (e.g., in layers like perception or ontology), an overtly (semi-)automatic implementation is not poised to produce an ontology-driven metadata model of requisite quality. To harness *the best of both worlds* (i.e., human and machine), the paper proposes a novel implementation of the conceptual disentanglement approach via Generative AI-driven Human-LLM Collaboration, wherein, the metadata librarian, using prompt engineering (Ekin 2023), exploits an LLM (Liang *et al*. 2022; Chang *et al*. 2024) to generate a conceptually (dis)entangled metadata model which, for each representational level, is validated/repaired/enriched by him/her. Notice that the validation or repair or enrichment of the Human-LLM collaboratively generated metadata model by the metadata librarian can involve several standard dimensions, e.g., metadata quality (Park 2009), semantic quality (Poels *et al*. 2005), and additionally purpose-specific dimensions such as, e.g., measuring the fit of the metadata model with actual cancer data/information. Overall, a widespread acceptance and implementation of the conceptual disentanglement approach via Generative AI-driven Human-LLM collaboration is poised to bring to the fore several advantages in terms of, e.g., metadata development methodologies, metadata crosswalk and interoperability (Khoo and Hall 2010), semantic heterogeneity (Hull 1997), FAIR data (Wilkinson *et al*. 2016), etc.

The remainder of the paper is organised as follows: The second section describes the different *knowledge representation levels* involved in designing a metadata model and how such levels are ordered and interlinked to cumulatively impact its design. The third section details the problem of conceptual entanglement individually and cumulatively across the representation levels and their *confounding impact* on the design of the metadata model. The Generative AI-driven Human-LLM collaboration based conceptual disentanglement approach individually and cumulatively across the representation levels and the way it minimises the entanglement and confusion in the design of the metadata model is elucidated in the fourth section. Finally, the fifth section discusses research implications from the related work in terms of some of the (generative) AI, metadata and semantics-based research issues academic libraries face and the sixth section concludes the paper. Throughout all the sections, especially for the second, third and fourth sections, an adapted instantiation of the motivating scenario in terms of a simplified ontology-driven metadata model for the *cancer domain* would be employed to exemplify and highlight the issues and approaches. Further, several of the fine-grained technical intricacies, e.g., of knowledge representation, are skipped in accordance with the broader methodological scope of the paper.

**Representation Levels in Modelling Metadata**

Let us now expand the discussion on the *characteristically independent but functionally interlinked* knowledge organization and representation levels which cumulatively compose any metadata model within the framework of an academic library. To that end, we have the following:

1. *Perceptual Level*, wherein, the correspondence between entities and their perception as concepts are modelled;
2. *Terminological Level*, wherein, the correspondence between perceived concepts and the terms which linguistically represent them are modelled;
3. *Ontological Level*, wherein, the correspondence between labelled concepts and their ontological commitment is modelled;
4. *Taxonomical Level*, wherein, the correspondence between ontologically enriched concepts and their classification into a taxonomic hierarchy is modelled;
5. *Intensional Level*, wherein, the correspondence between the taxonomic concepts and how they are intensionally interrelated and described is modelled.

Additionally, it is also important to note that the motivating example considered for illustrating the above (and for remainder of the paper) are three equally probable variations of an ontology-driven library metadata model for cancer domain (e.g., encoding concepts such as Clinical Trial, Patient, Biomarker, Imaging Test, Histopathology Report, etc.) generated by prompting the Generative-AI based LLM (interface) - ChatGPT 3.5 - using a series of prompts. Please note two observations (valid throughout the paper). The metadata models cannot be reproduced completely within the full-text of this paper due to constraints of space and to that end, please follow the link - motivating example - wherein readers can get a fuller understanding of the concept/property exemplified in the paper and its page location within the linked document describing the generated metadata models. Also, while the metadata model generated by ChatGPT 3.5 is informal in nature, this is not relevant to the scope of this paper as it can be directly formalised in any formal language of choice (e.g., the web ontology language - OWL) for concrete implementation purposes. The above levels are elucidated as follows.

First, let us concentrate on the *perceptual level* which, while being *not explicitly pronounced* in state-of-the-art library metadata research and implementation (see, e.g., (Haynes 2018; Strecker *et al*. 2021)), is nonetheless the very first level where the crucial representational choice as to how a set of entities should be perceived as concepts (later) composing a library metadata model is made. Notice two key dimensions which inform the decision-making of a metadata librarian at this level. First, it goes without mentioning that perception is highly *egocentric* to an individual academic library user and therefore is *usually incomplete* (Bagchi 2021a) and cannot be fully captured in a (semi-)formal manner. Second, while the first dimension holds, it is also equally the case that various *communities of practice* (Wenger 1999; Cox 2005) (e.g., users of different types of libraries) commit to a shared and homogeneous perception as to how certain entities should be perceived as concepts. Consider the case of the motivating example. In it, pages 1-7 elaborates a general (ontology-driven) library metadata model for cancer domain generated by ChatGPT 3.5 with constituent concepts (classes and properties) at three levels of abstraction. Further, note how the general model, via successive prompts, is tuned to the potential perceptions of a representative library user belonging to the specialised cancer research community (pages 8-10), oncology policy community (pages 10-12) and medical college student community (pages 12-14), respectively. It is interesting to observe that even for the same macro domain (i.e., cancer), each of the three metadata models encode concepts and properties *uniquely relevant to the perception* of a specific user community.

Second, given the understanding as to why perception is crucial for modelling metadata, let us now turn to the *terminological level* where the key representational choice is to decide how a set of perceived concepts should be labelled using a body of terminology. This choice is *not trivial* for the metadata librarian due to the very nature of the interaction of language with the perceptual level. Languages are *"itemized [terminological] inventories"* (Brown and Lenneberg 1954) of the entities we perceive and each of such *"inventories"* generate either *similar (but not the same)* or a different lexicalization of the perception due to issues following from *linguistic relativity* (Boroditsky 2011). Such a lexicalization might manifest amongst different sets of academic library users in terms of different linguistic phenomenon. For example, two well known linguistic phenomenon include *polysemy* (the coexistence of several potential meanings of a term) and *synonymy* (terms having similar meanings) (Glynn and Robinson 2014). *Lexical gaps* (Bentivogli and Pianta 2000) occur when, for a concept, a specific language/vocabulary doesn't have a referent term. The scenario might get even more confounded when library user communities communicate with each other or with the computer system by linguistically referring to the perceived entities with terms from specialised terminological standards (Suonuuti 1997) or glossaries (Sarginson *et al*. 2012) possibly from different languages. There might be issues with both mapping and interoperability of terms amongst the standards due to issues of linguistic phenomenon or non-existent terminological crosswalks. To exemplify, in the motivating example, please refer to page 20-23 which detail the LLM response to the prompt clarifying the terminology employed in the three examples. Notice that parts of the cancer research library metadata model uses terms from the Dublin Core (Weibel *et al*. 1998) to express its perceived concepts. Similarly, the oncology policy library and the medical college library metadata models use terms from standards such as HL7 (Health Level Seven) standard, National Comprehensive Cancer Network (NCCN) standard, SNOMED-CT, etc. (Schulz, Stegwee and Chronaki 2019).

Third, given the central importance of terminology in the design of any library metadata model, the next dimension to understand is the *ontological level* where the key representational choice regarding the ontological commitment (Guarino, Carrara and Giaretta 1994) of the (newly) labelled concepts is made. This choice, again, is non-trivial for the metadata librarian due to the theoretical root of the interaction of ontology with the terminological level, in the sense that, committing to a terminology results in committing to an ontology (Moltmann 2019) which might not be necessarily completely compatible with another ontology/terminology pairing. The chief importance of this representation level is to explicate the otherwise implicit commitment of the body of terms to a *top-level ontology*

(Guarino 1997), wherein, such an ontology, based on an explicitly specified philosophical doctrine, reveals the underlying nature of domain-level concepts. There can be several such philosophical doctrines such as three-dimensionalism, four-dimensionalism, etc. (McCall and Lowe 2006), and according to the chosen doctrine, the domain-level labelled concepts can be classified as a kind, a part-of, an event, a process, a role, a function or a property (Gangemi *et al*. 2001). Two observations. First, by far, this representation layer is the *most demanding* in terms of the involvement of the human modeller, i.e., the metadata librarian, for the simple reason that while different library user communities might, most possibly, conceptualise the same set of entities into different top-level ontological categories, it is chiefly the responsibility of the metadata librarian to explicate such assumptions. Secondly, as evidenced in decades worth of research in semantic heterogeneity (Hull 1997), the ontological level is key to achieve a range of semantics-intensive tasks like semantic mapping (Beneventano *et al*. 2008) and semantic interoperability (Bittner, Donnelly and Winter 2005) with respect to (meta)data models. To exemplify, in the motivating example, please refer to page 23-29 which detail the LLM response to the prompt requesting for the ontological categories employed in the three examples. For example, notice that in page 26, the LLM response aligns concepts and properties such as Document and hasAuthor to top-level ontological categories like Information Object and Quality.

Fourth, given the importance of ontological commitment of constituent concepts and properties in a library metadata model, let us concentrate on the *taxonomical level* where the key representational choice regarding the ontologically characterised concepts and their classification into a taxonomical hierarchy is made. This choice by the metadata librarian is guided by two key factors. First, the ontological level already induces a very abstract high-level taxonomy for the concepts due to the mandatory ontological constraints (Guarino, Carrara and Giaretta 1994) they pose while the metadata librarian commits to a specific top-level ontology and its constituent top-level ontological categories. For example, if the concept Person and Patient belong to the top-level ontological categories Kind and Role, it is an ontological constraint that Role can never be the taxonomical parent of a Kind, and, therefore, Patient can never be the taxonomical parent of Person. Within the ontological constraints, however, the taxonomy can be designed by the metadata librarian in accordance with how its hierarchy and metadata properties will be exploited by the target information service (e.g., a data catalog (Guptill 1999), a chatbot (Bagchi 2020)). To exemplify, in the motivating example, please refer to page 30-34 which detail the LLM response to the prompt requesting for the taxonomical choices made (for concepts as well as properties). For example, in page 30, Library Resource is specialised into Research Paper and Dataset. Notice that, alongside the ontological level, the taxonomical level is also *equally demanding*, if not more, in terms of the involvement of the metadata librarian in modelling the classification hierarchy.

Last but not the least, given the key importance of the taxonomic classification in the composition of a library metadata model, let us focus on the *intensional level* where the key representational choice is how to interrelate and describe each individual concept in the taxonomy. This is done by first finalising the specialisation of the already classified properties (in the ontological level) into object properties and data properties (Bagchi and Madalli 2019). Note that, even if the properties are informally classified as object or data property by the metadata librarian before this level, the final decision is taken at this level. Second, each concept in the taxonomy is assigned with a set of object properties which encode how the concept is interlinked with other concepts in the overall metadata model. Equally, and perhaps the most important aspect, every taxonomic concept is assigned with a set of data properties which describe its attributes. Two observations. First, this level, implementationally, is perhaps the most visible and pronounced level in a library metadata model as it decides on the data properties which ultimately encode real-world library metadata. Secondly, this level is central to a library metadata model in the sense that it decides the inheritance of data properties by classes (aka concepts) in the taxonomy, thereby, considerably influencing the decision concerning the description of each taxonomic concept by a set of data properties and data types. To exemplify, in the motivating example, please refer to page 34-39 which detail the LLM response to the prompt requesting for the object and data property choices made. For example, in page 35, the concept Policy Document is described with data properties like *{policyDocumentID, policyDocumentTitle, policyDocumentText}*.

Finally, it is interesting to note two crucial highlights about the aforementioned levels of knowledge organization and representation which constitute any library (or, even a generic) metadata model. First, the ordering from perception to intensionality is not just a linear, ordered sequence but is essentially an ordered and continual *interlinked scientific spiral* (Ranganathan 1957) from perception to intensionality and back. To that end, successive cycles can be utilised by the metadata librarian to continuously validate/repair/update/enrich the library metadata model with new potential changes in any of the representation levels. This also *reinforces* the notion of the *non-validity* of a single or a few metadata models as necessary and sufficient for any and every domain and application scenario. Second, notice that the above levels of representation is a general framework to understand the composition of a library metadata model. It might well be the case that for a specific application scenario, the perceptual level or the terminological level or both are quite homogeneous and, therefore, of much less relative importance to the model than that of, e.g., the ontological level or the intensional level. In any case, such *permutations and combinations* of the relative importance of each representation level to the final library metadata model should be decided on a *case-by-case* basis.

**Entanglement in Modelling Metadata**
Given the central and irreplaceable impact of the five functionally interlinked representation levels on the modelling of metadata, let us now focus on how the conceptual entanglement problem is *implicitly instantiated* at each successive level within the metadata design strategy of an academic library. To that end, we have the following:
1. *Perceptual Entanglement*, referring to the representational manifoldness between entities and their perception as concepts;
2. *Terminological Entanglement*, referring to the representational manifoldness between perceived concepts and the terms which are employed to label them;

3. *Ontological Entanglement*, referring to the representational manifoldness between labelled concepts and their ontological commitment;
4. *Taxonomical Entanglement*, referring to the representational manifoldness between ontologically enriched concepts and their taxonomic classification;
5. *Intensional Entanglement*, referring to the representational manifoldness between the taxonomic concepts and their intensionality.

In order to exemplify the stratification above, the same motivating example of the ontology-driven library metadata model for cancer domain is employed (pages 8-43). Notice also the fact that the stratification of conceptual entanglement proposed above is in *sync* with the stratification of representation in metadata proposed in the second section. The above levels are elucidated as follows.

First, let us concentrate on perceptual entanglement which, in essence, refers to the many-to-many mapping between entities and their perception as concepts which would eventually compose a library metadata model. There are two key factors which underlie the entanglement in the decision-making of a metadata librarian at this level. First, as already briefed before, perception is egocentric and can be unique to communities of practice. This premise leads to the notion of *perception as a cognitive filter* (Guarino, Guizzardi and Mylopoulos 2020), i.e., in our terms, the fact that the same entity and its properties can be perceived differently by different groups of library users depending on their overall purpose and goals. There can be an overlap in the concepts perceived between, say, two groups of library users and, equally, there can be ranges of mutual exclusion in their perception (e.g., due to different purposes or goals). Second, the highly implicit nature and consideration of perception within academic library metadata design (see, e.g., (Fleming, Mering and Wolfe 2008; Lopatin 2010)) almost leads to the necessary and sufficient assumption (explained before) on surface while the underlying multiplicities of perception remain unaddressed. Consider the case of the motivating example generated by ChatGPT 3.5 via Human-LLM collaboration. Three different user perceptions, *viz.,* of a representative cancer research library (pages 8-10), an oncology policy library (pages 10-12) and a medical college library (pages 12-14), of the same macro domain cancer are illustrated in the example. Notice that the commonality in the perception of the three different communities are indicated by their shared perceived concept of cancer specialised by its various types (breast cancer, lung cancer, etc.) and described by common properties (e.g., biomarkers associated with a cancer). However, major differences emerge in the three perceptions due to their very purpose in studying the same entity cancer. The cancer research library user community is interested in concepts like research study, treatment and properties like clinical trial ID, pathology report text, etc. The oncology policy library user community, on the other hand, is interested in policy documents like clinical guidelines, best practice, regulatory policy, etc. The medical college library user community is interested in medical study, study material, treatment approach, etc. Please see pages 15-18 in the motivating example for more details on differences in perception. It is interesting to notice from above the representational manifoldness between the same (domain) entities and how they are variously perceived as concepts.

Second, let us focus on the terminological entanglement which implies the necessary existence of a many-to-many mapping between perceived concepts and their lexicalization using a terminological label. As briefly mentioned earlier, linguistic phenomena are a *key inducer* of representational manifoldness at this level. For example, polysemous terms labelling a perceived concept provides at best an ambiguous notion of its meaning and magnifies the many-to-many possibilities for the metadata librarian. On the other hand, the central problem with synonyms is the establishment of mapping. Given implicitly synonymous terms, it might not always be straightforward for the metadata librarian to infer whether they are same or synonymous or a broader/narrower/distinct term. The above problems are further compounded in settings which might involve multiple languages and/or multiple communities of different genres of academic library users. To exemplify, in the motivating example, let us consider the singular case of the concept Breast Cancer (page 8). It can be termed variously by different communities of library users as Carcinoma of the Breast, Mammary Carcinoma, Breast Tissue Malignant Neoplasm or Lobular Adenocarcinoma. In multilingual metadata modelling settings, the same concept can be referred to as, for instance, *cancer du sein* in French, *cáncer de mama* in Catalan or *rakovina prsu* in Czech. While technologies from Natural Language Processing (NLP) can be harnessed to implement some of these semantic (non)equivalences, it is crucial for the metadata librarian to first diagnose and understand the entanglement and manifoldness existent in the equivalency mapping.

Third, given the elucidation of terminological entanglement, let us now elaborate the notion of ontological entanglement which refers to the necessary existence of a many-to-many mapping between labelled concepts and their ontological commitment. There are two interlinked dimensions underlying the above entanglement. First, due to perceptual and subsequent terminological entanglement, different communities of library users can perceive and label the same entity differently, thereby, bootstrapping the many-to-many mapping between different entities/concepts and the different top-level ontological categories they might potentially be categorised into. Second, as a consequence of the first reason, different communities of library users, unknowingly and implicitly, commit to the philosophical doctrine of the one of the state-of-the-art top-level ontologies such as DOLCE (Gangemi *et al*. 2002), BFO (Smith, Kumar and Bittner 2005), UFO (Guizzardi *et al*. 2022), etc., thereby, adding a second layer of many-to-many entanglement between entities and top-level ontological theories. The ontological entanglement resulting in a *multiplicity of representational manifoldness* between labelled concepts and their top-level ontological characterization, in effect, translates into application-level implementational entanglements such as the management of interoperability and harvesting of metadata in networked academic library settings (Taha 2012). To exemplify, in the motivating example, let us consider the ontological entanglement in the oncology policy library metadata model as detailed in page 27. Notice that the same concept of a Policy Document has been categorised into different ontological categories (Information Object, Artefact, Continuant) by different top-level ontologies (DOLCE, UFO, BFO), respectively. Further, each of these categories can be instantiated for various other labelled concepts, e.g., textbook on page 28 (having a different semantics).

Fourth, with the assumption of ontological entanglement and ontological constraints, let us now elucidate the taxonomical entanglement sub-problem which refers to the necessary existence of a many-to-many mapping between ontologically enriched concepts and their taxonomic classification which would eventually constitute the backbone of a library metadata model. There are four key parameters, from Ranganathan's *faceted classification* theory (Ranganathan 1967; Ranganathan 1989), which induce representational manifoldness at this stage. First, with respect to a concept at a specific level of abstraction in the taxonomy, there are always multiple characteristics which can be employed to taxonomically specialise that concept into (potentially many) subordinate concepts. Second, the successive application of characteristics across the entire depth of taxonomy (with the possibility of multiple classificatory characteristics at each level of abstraction) leads to potentially infinite entangled classifications. Third and fourth, there can also be multiple ways in which concepts can be organised horizontally across a specific level of taxonomic abstraction (termed *arrays* in (Ranganathan 1967)) and vertically across a taxonomic path (termed *chains* in (Ranganathan 1967)), respectively. To exemplify the above, in the motivating example, please refer to page 30-34 which detail the LLM response to the prompt requesting for the taxonomical choices made. Although simplistic, note that the Human-LLM prompt collaboration returns two equally probable but different entangled taxonomies (page 31: example 1 and example 2) for the oncology policy library user community. It is interesting to observe that in example 1, the taxonomy is implicitly suited for a library metadata model on social-healthcare policy perspective whereas in example 2, the taxonomy is implicitly suited for a library metadata model on governmental-healthcare policy perspective.

Last but not the least, with the assumption of taxonomical entanglement, let us now elucidate the problem of intensional entanglement which refers to the necessary existence of a many-to-many mapping between the taxonomic concepts and their intensional interrelation and description. There are two key factors which magnify representational manifoldness in this final stage. First, during the final decision to characterise a property (uncovered at the ontological level) as an object or a data property, there exists a many-to-many mapping as the same property can be represented and refactored as an object or a data property depending on the purposes and goals to be served by the library metadata model. Second, a concept in the taxonomy can be interrelated and described via multiple possible combinations of sets of object and data properties, again, depending on the precise purpose of the final metadata model. To exemplify the entanglement, in the motivating example, please refer to page 34-39 which detail the LLM response to the prompt requesting for the object and data property choices made. For example, the Human-LLM collaboration example generated two equally probable but different sets of object and data properties (page 36) implicitly suited for the medical students and medical study participants user community of a medical college library, respectively.

Finally, notice that the aforementioned individual *level-by-level* representational manifoldness are *accumulative* in nature and, therefore, there is also an overall manifoldness between entities and their taxonomy and intensional characterization. In the motivating example, pages 39-43 detail at length the LLM response to the prompt requesting for the overall many-to-many mapping between entities, taxonomical hierarchies and intensional property-based characterization. For example, even within the same macro domain of discourse cancer, different perception by the three different communities of library users (cancer specialists, oncology policy experts and medical college professionals) generate multiple terminology, ontological categories, taxonomies and property predication, most of which are not in correspondence amongst each other. Further, as also previously noted in the second section, it might well be the case that for an application scenario, there might be no manifoldness in perception or in terminology or both (with, e.g., the other entanglements still holding). In any case, the conceptual entanglement problem is a generic characterization inherent and implicit by design in any library metadata model and should be adapted as necessary on a case-by-case basis.

**Towards Generative AI-driven Disentangled Metadata Modelling**
After the elucidation of how the conceptual entanglement problem is inherent, by design, at each representation level, let us now describe in detail how a metadata librarian can exploit the conceptual disentanglement approach, via *Generative AI-driven Human-LLM collaboration*, to disentangle entangled metadata within the framework of an academic library. To that end, we have the following:
1. *Perceptual Disentanglement*, referring to the explicit representational bijection between entities and their perception as concepts;
2. *Terminological Disentanglement*, referring to the explicit representational bijection between perceived concepts and the terminology used to linguistically label them;
3. *Ontological Disentanglement*, referring to the explicit representational bijection between labelled concepts and their ontological commitment;
4. *Taxonomical Disentanglement*, referring to the explicit representational bijection between ontologically characterised concepts and their precise taxonomy;
5. *Intensional Disentanglement*, referring to the explicit representational bijection between the taxonomic concepts and their intensionality.

In order to exemplify the stratification above, the same motivating example of the ontology-driven library metadata model for cancer domain is employed (pages [8-43](#)). Notice also the fact that the stratification of conceptual disentanglement proposed above is in *sync* with both the stratification of conceptual entanglement and the stratification of representation in metadata proposed in the second and the third section, respectively. The above levels are elucidated as follows.

First, let us concentrate on the perceptual disentanglement sub-approach which advocates, on the part of the metadata librarian (i.e., the modeller), to decide on an explicit *one-to-one* mapping between entities and their perception as concepts. As described and exemplified in the third section, even for the same (domain) entities, there is an inherent representational manifoldness which entangles the resultant library metadata model. There are two key factors which underlie the disentanglement approach of a metadata librarian at this level. First, the metadata librarian has to generate, via prompt engineering (similar to that in the motivating example), an initial characterization of the perception he/she wants to follow in designing the metadata model. Notice that the goals and purposes which such a model would potentially serve is a key influence on the design of prompts at this stage to elicit a reasonable response from the LLM. Given an initial record of perception generated via the above Human-LLM collaboration, the next crucial step for the metadata librarian is to employ (a combination of) several standard techniques to validate and consolidate the generated perception depending on the *criticality* of the use of the metadata model. To that end, he/she can conduct a lightweight digital ethnographic exercise (Varis 2015) about the user community and its perception about a domain. He/she can also employ (digital) focus groups (Morgan 1996) and/or one-to-one consultation with domain experts (Drabenstott 2003) to better understand a community's viewpoint about entities of a specific domain. Further, independently or in addition to the above, the metadata librarian can also perform an information and data validation exercise, wherein, he/she can explore real-life information and data resources produce by an user community to gauge its perception about a domain or even, if available, reuse existing conceptual disentanglement documentations (thereby, reinforcing the spiral nature of the levels). Finally, the above inputs can be consolidated to validate/repair/enrich the initial perception and, hence, facilitate disentangling the perceptual entanglement. To exemplify, in the motivating example, the metadata librarian, after the perception validation exercise, might choose to add a new concept multi-omics workflow and related properties to the perception already generated by the Human-LLM collaboration for cancer research library user community. Notice also the fact that the initial perception generated via the LLM prompting also *reduces* to a significant extent the otherwise laborious exercise of domain analysis (Hjørland and Albrechtsen 1995) on part of the metadata librarian.

Second, assuming the representational bijection at the perceptual level, let us now focus on terminological disentanglement, which advocates, on the part of the metadata librarian, to decide on an explicit one-to-one mapping between perceived concepts and the terminology which would compose the disentangled library metadata model. The guiding cardinal which a metadata librarian can employ for disentanglement at the terminology level can be referred to as the principle of *user warrant*. It is based on the generalised notion of *warrant* in information science (Nylund 2020; Lancaster 1972) and can be understood as the principle to assign to a perceived concept a linguistically disambiguated and semantically explicit term which is based on the documented warrant of the (expert) library users. To that end, each (meta)data concept/property term label should embody standard terminological quality, e.g., having a natural language gloss, examples, identifiers, etc (Miller 1995). There are several (combinations of) options which a metadata librarian can exploit to achieve representational bijection of terminology. If the user warrant is to use relatively commonsense terms to refer to concepts of cancer in a, e.g., oncology policy library, lexical-semantic resources such as WordNet (Miller 1995) might be exploited to achieve the representational bijection. On the other hand, if the user warrant in a specialised cancer research library is to refer to concepts of cancer using scientific terms, resources such as specialised glossaries and healthcare/clinical terminological standards (Schulz, Stegwee and Chronaki 2019) can be exploited to achieve an explicit one-to-one mapping. Further, in multilingual settings, NLP resources like multilingual variations of WordNets and WordNet-like resources (Pianta, Bentivogli and Girardi 2002; Dash, Bhattacharyya and Pawar 2017; Navigli and Ponzetto 2012) and/or open multilingual knowledge bases like Wikidata (Vrandečić and Krötzsch 2014) can be exploited to disentangle and disambiguate terms.

Third, assuming the representational bijection at the terminological level, let us move to the next activity of ontological disentanglement, according to which, the metadata librarian has to decide on an explicit one-to-one mapping between labelled concepts and their ontological commitment. Notice that, similar to the disentanglement strategy at the terminological level, the *general guidance* of user warrant is key to uncover the ontological commitment of library users at this level. To that end, the metadata librarian can apply a four-staged approach. First, the librarian can prompt an LLM like ChatGPT 3.5 appropriately (see, for instance, the examples in pages 23-29 of the motivating exam- ple) to bootstrap an initial 'entangled' version of the alignment of the library metadata model with different state-of-the-art top-level ontologies. Second, the librarian can reuse the outputs and documentation of focus group interviews and/or domain expert consultation and/or relevant conceptual disentanglement documentations (as discussed before in the perceptual disentanglement level) to understand the ontological warrant of the community of library users he/she is attending to. Third, he/she uses the results of the second step to explicitly choose and enrich the best possible ontological fit amongst the different ontology alignments produced in the first step. Finally, the metadata librarian also should explicate the ontological categories of each individual labelled concept and property that is under consideration with respect to the chosen top-level ontology.

Fourth, assuming the representational bijection at the ontological level, let us now concentrate on taxonomical disentanglement, which advocates, on the part of the metadata librarian, to decide on an explicit one-to-one mapping between ontologically characterised concepts and their precise taxonomy. Given the general direction from the needs of the library and information service(s) which would exploit the final library metadata model, the metadata librarian can follow a four-step disentanglement strategy in response to the four stages of taxonomical entanglement (see the third section). The solution is grounded in the *canons of knowledge classification* proposed by Ranganathan in his faceted classification theory (Ranganathan 1967; Ranganathan 1989. First, the canons of *characteristics* (Ranganathan 1967) like canons of relevance and ascertainability should be applied to eliminate manifoldness at the level of selecting a classificatory characteristic for specialising concepts at a specific level of abstraction in the taxonomy. Second, the *canons of succession of characteristics* (Ranganathan 1967) like canon of relevant succession should be applied to disentangle the multiplicity existent in how classificatory characteristics are successively applied to design the conceptual depth of the taxonomy. Third and fourth, the *canons of arrays and chains* (Ranganathan 1967) should be applied to disentangle the manifoldness existent while modelling concepts across a specific horizontal level and across a path of the taxonomy, respectively. To aid the above exercise, the metadata librarian can also consult and reuse relevant state-of-the-art taxonomies if available as open source code (Bagchi 2018). Finally, the above procedure is expected to generate an enriched and disentangled version of one of the many possibilities initially generated via Human-LLM collaboration, e.g., as exemplified in the motivating example (page 30-34).

Last but not the least, assuming the representational bijection at the taxonomical level, let us move to the final activity of intensional disentanglement, according to which, the metadata librarian has to decide on an explicit one-to-one mapping between the taxonomic concepts and their intensional interrelation and description. To that end, he/she can proceed on a two-step strategy. First, he/she should make explicit a final decision on the exact *split* of properties into two distinct sets: a set of object properties and a set of data properties. This decision is critically influenced by the information service applications driven by the warrant of a specific community of library users. Given the first step, the next step is to make explicit the decisions on exactly how the *object properties* would be exploited to interlink the concepts in the disentangled taxonomy, i.e., determining the precise conceptual domain and ranges of the object properties. Further, perhaps most importantly, he/she should determine the exact set of *data properties* which should describe each concept in the taxonomy (also factoring in the taxonomic inheritance). Together, the above steps can facilitate a representational bijection and, thereby, a generate the final disentangled ontology-driven library metadata model out of the many intensionally entangled possibilities initially generated via Generative AI-driven Human-LLM collaboration, e.g., as exemplified in the motivating example (page 34-39).

There are three key observations with respect to the aforementioned elucidation of the levels of conceptual disentanglement. First, as with the second and the third section, it might well be the case that for a specific application scenario, the conceptual disentanglement exercise is relevant to be implemented only for specific representation levels and not for all five levels together. Second, notice that the disentanglement at each of the above levels is effectuated by Human-LLM collaboration via appropriate prompt engineering. It is definitely the case that in some levels, e.g., ontological and taxonomical disentanglement, human modellers should be more involved than LLMs due to the very *conceptual and semantics-intensive nature* of disentanglement. On the other hand, the initial responses generated by LLMs can speed up the bootstrapping of modelling metadata by several counts. Notice also the fact that *human oversight*, even if not demanding, is crucial, at each level, to minimise the effect of the so-called *LLM hallucinations* (Rawte, Sheth and Das 2023; Chomsky, Roberts and Watamull 2023). Finally, the conceptual disentanglement exercise should be formally documented as a *datasheet* (Gebru *et al*. 2021) (for possible future reuse) with all the decisions concerning the disentanglement of the manifoldness into one-to-one correspondences clearly explicated.

**Research Implications for Libraries and Beyond**
The deconstruction of a library metadata model as composed of five disentangled representation levels is a *novel way of thinking* about library metadata and is directly linked to some of the AI, metadata and semantics-based research issues which academic libraries are increasingly grappling with. To that end, let us focus on general perspectives relevant to related work in library metadata and library knowledge discovery, FAIR data and metadata, and AI, *Generative AI and academic libraries* vis-a-vis the conceptual entanglement and disentanglement metadata modelling proposal advanced in the current work. Each of the perspectives are individually discussed as follows.

Knowledge Discovery, according to the highly cited work in (Fayyad, Piatetsky-Shapiro and Smyth 1996), can be defined as *"the non-trivial process of identifying valid, novel, potentially useful and ultimately understandable patterns or knowledge in data"*. Notice that, in the above quoted definition, what is assumed *a priori* is the notion of user warrant based data modelling and data annotation without which eliciting patterns from such data would be an uphill task. In fact, some of the assumptions (alongside many others) were alluded to in later works such as (Pazzani 2000). Notice that metadata-based data modelling and annotation is key to digital infrastructures, e.g., digital libraries (Borgman 1999), discovery services (Vaughan 2011), etc., of academic libraries which host and expose key assets for the academic library user community, especially the community of data science researchers, to engage in algorithmic knowledge discovery. To that end, the mainstream research in library metadata (see, e.g., (Dunsire and Willer 2010; Dunsire and Willer 2011; Coyle 2015; Peponakis 2013)) have overwhelmingly concentrated on developing metadata schemas (which is key) but have, at best, *under-considered* the key spectrum of *human factors in modelling metadata*. This has led to long standing issues with knowledge discovery in academic libraries (Richardson, Srinivasan and Fox 2008), metadata crosswalks (Khoo and Hall 2010), metadata and data interoperability (Pagano, Candella and Castelli 2013), etc. More recently, the CILIP research report (Cox 2021) suggested enhanced emphasis on the *human and social aspect* and termed knowledge discovery as a *socio-technical* process. The proposal of Generative AI-driven conceptual disentanglement proposed in this work across the five representation levels right from (human) perception to the formalisation of a library metadata model can contribute as a *research catalyst* to this under-researched theme. Moreover, the level-by-level disentanglement of a library metadata model can form the base for developing incremental frameworks and methodologies for attenuating the impact of, e.g., metadata non-interoperability and skewed crosswalks.

Next, let us focus on how the current work is directly linked to the upsurge of research in *FAIR* data (Wilkinson *et al*. 2016), i.e., Findable, Accessible, Interoperable and Reusable data. As stated in its original vision (Wilkinson *et al*. 2016), the effort to make data (or, in general, an information resource) FAIR involves *a dedicated metadata strategy* on part of the implementing entity. Academic libraries and the library metadata community have been quick to incorporate the FAIR vision (especially the Findability and Accessibility dimensions) and have been successful in the technological infrastructure (e.g., data catalogues) backing a standalone FAIR implementation (for example, see, (Hettne *et al*. 2020; Koster and Woutersen-Windhouwer 2018)). However, as the current proposal of conceptual disentanglement shows, FAIRification is a multi-level strategy encompassing all the five representation levels: perception, terminology, ontology, taxonomy and intensional characterization. This is key, especially for the Interoperability and Reusability dimensions as there can be non-interoperability in each of the five representation levels due to representational manifoldness leading to non-reusability of digital data and information (thereby, also defeating the overall purpose of FAIR). Additionally, the above aspect also reinforces the importance of ontology-driven metadata modelling as, without an ontology structuring and consolidating the (meta)data properties, it is next to impossible, as per the current technology toolbox, to achieve interoperability and reusability both for data and ontology (see, e.g., (Fernández-López *et al*. 2019; Carriero *et al*. 2020) for related issues and problems). Finally, while the conceptual disentanglement framework is innovative in its vision of level-by-level FAIRification, it runs the risk of overwhelming intellectual work on part of, e.g., a metadata librarian. The Generative AI-driven Human-LLM collaboration strategy to achieve conceptual disentanglement via prompt engineering is advanced exactly to mitigate a substantial percentage of initial work which can be automated and the human modeller can validate and enrich it with his/her expertise, differently in different representation levels.

Finally, let us also move to a more abstract plane and position the current work in terms of how AI in general and *Generative AI* in particular have impacted academic librarians, information professionals and users. The CILIP report (Cox 2021), as cited before, provides a well-rounded bird's-eye view as to how AI is going to shape the academic library space in the years to come. The possibilities of generative AI applications in (academic) libraries encompasses a rich spectrum of information-intensive activities including, amongst others, intelligent web and mobile search, socio-technical infrastructures for knowledge discovery, chatbots, marketing and user behaviour elicitation, robotics, data science community building and data stewardship. Notice three dimensions which are key to the *socio-technical tool-box* required to conceptualise and implement all of the above activities. First, all of the above activities depend on the *involvement of humans*, particularly the warrant of library users but also the technical expertise and knowledge of academic librarians and other information professionals. Second, all of the above activities also depend on the *involvement of machines*, i.e., computers and computing infrastructures to semi-autonomously implement digital information services and increase their usage penetration amongst users. Thirdly, all of the above activities crucially depend on *metadata*, which forms the *glue* between *humans* (helping them discover and retrieve data, information and knowledge from machines) and *machines* (to exploit and fine-tune user warrant, services and information discoverability). In this context, the current proposal of a conceptual disentanglement strategy for modelling library metadata via Generative AI-driven Human-LLM collaboration contributes to all the three dimensions. The strategy involves humans (metadata librarians as semantic modellers and users), machines (generative AI technology (Sætra 2023; Jo 2023; Fui-Hoon Nah *et al.* 2023) in the form of LLMs) and a strategy to model disentangled metadata via greater Human-AI collaboration.

**Conclusion and Future Work**
To summarise, the paper presented a novel and completely general Generative AI-driven Human-LLM collaboration approach to model disentangled library metadata free from several levels of representational entanglements which can otherwise have unintended technological consequences. To that end, the paper advanced three key contributions. First, it introduced a novel way of thinking about library metadata in terms of a composition of five functionally interlinked representation levels from perception to intensional definition. Second, it introduced the implicit representational manifoldness existent in each of the five levels which cumulatively contribute to a conceptually entangled library metadata model. Third, it proposed a Generative AI-driven Human-LLM collaboration based conceptual disentanglement approach to disentangle the entanglement inherent, level-by-level, in the metadata model leading to a disentangled metadata model which is validated, explanatory and continually enriched via a spiral process. Some of the future research lines in this direction can include a new conceptually disentangled interpretation of FAIR (meta)data in libraries, elevating the approach to a full-fledged general metadata modelling methodology beyond libraries and a detailed cost-benefit analysis of the Human-LLM collaboration dimension in conceptual disentanglement.

**Author Bio**

Mayukh Bagchi is a postdoctoral researcher specializing in Information Science and Artificial Intelligence at the Department of Information Engineering and Computer Science (DISI) at the University of Trento, Italy, where he is also a Graduate Teaching Assistant in the Master of AI and Data Science course on Knowledge Graph Engineering from 2020. He is also an affiliated researcher at the Institute for Globally Distributed Open Research and Education (IGDORE). He has graduate degrees in Mathematics, Computer Science and Information Science. He has participated in several international AI-based research projects like JIDEP (Horizon Europe Research and Innovation Project), MIUR-Italy DELPHI Project and DataScientia Data Space Research Project. His research interests are in Interdisciplinary AI, Interdisciplinary Information Science, Human-AI Interaction, Philosophy of Machine and Deep Learning, Foundations of Mathematics and Literary Epistemology.